\documentclass[pdflatex,sn-mathphys-num]{sn-jnl}

\usepackage{graphicx}%
\usepackage{multirow}%
\usepackage{amsmath,amssymb,amsfonts}%
\usepackage{amsthm}%
\usepackage{mathrsfs}%
\usepackage[title]{appendix}%
\usepackage{xcolor}%
\usepackage{textcomp}%
\usepackage{manyfoot}%
\usepackage{booktabs}%
\usepackage{algorithm,caption}%
\usepackage{algorithmicx}%
\usepackage{algpseudocode}%
\usepackage{listings}%
\usepackage{bbm}%

\theoremstyle{thmstyleone}%
\theoremstyle{thmstyletwo}%

\theoremstyle{thmstylethree}%

\begin{document}

\title[Article Title]{Optimal Inference of Asynchronous Boolean Networks}


\author[1]{\fnm{Guy} \sur{Karlebach}}\email{gkarleba@fitchburgstate.edu}

\affil[1]{\orgdiv{Department of Computer Science}, \orgname{Fitchburg State University}, \orgaddress{\street{Pearl Street}, \city{Fitchburg}, \postcode{01420}, \state{MA}, \country{USA}}}


\abstract{The network inference problem arises in biological research when one wishes to explain a phenotype using a network of interactions between molecules.  The diverse nature of the data and nonlinear dynamics of the network pose significant challenges in choosing the best model.  In addition to balancing fit and model size, computational efficiency must be considered.  The latter constitutes a central consideration for the researcher since  underlying the measurements, which are affected by experimental noise, there is a complex computational mechanism that may be asynchronous and is inherently hard to identify.  To address these challenges, we present a novel approach that uses algorithmic complexity to infer an asynchronous Boolean network model from experimental data.  We present an algorithm that is optimal within this framework and allows for asynchronicity in network dynamics.  Results are described for real data, a literature-derived network and random networks.}




\maketitle

\section{Introduction}\label{sec1}
Research in molecular biology often aims to reveal a mechanistic understanding of underlying processes, in contrast to correlations between observations.  In particular, one aims to describe cellular networks of interactions between proteins and other molecules that generate a specific phenotype \cite{Karlebach2008,Arenas2015,Hackett2016,Eichenberger2004}.  Often, these networks display complex, nonlinear dynamics due to combinatorial interactions between regulators and feedback loops in the topology. \cite{Kauffman1969}.  A key aspect of the problem is the Curse of Dimensionality, which ensures that the number of networks and datasets grows exponentially with the number of involved proteins, making inference susceptible to overfitting.
Gene regulatory networks, which control the expression of genes in the cell, are an important type of biological networks that are studied using transcriptomic technologies. These technologies measure the level (expression) of all the genes in a sample and, more recently, in single cells \cite{Andrews2020}, allowing the detection of regulatory relationships between them.   When the measurement is at the sample level, we refer to the data as RNA-Sequencing data (RNA-Seq), and when it is at the single-cell level as single-cell RNA-Seq (scRNA-Seq).
Numerous, fundamentally different inference methodologies have been proposed.  For example, Keyl et al. used an explainable AI approach and layer-wise relevance propagation in order to select gene regulators that are ranked as important predictors in a neural network model \cite{Keyl2023}.  SCODE is a method for inferring ordinary differential equations from scRNA-Seq data, such that the regulatory relationships can be derived from the equations \cite{Matsumoto2017}.  CEFCON uses a graph neural network with an attention mechanism for constructing the regulatory network from prior network knowledge, scRNA-Seq data and trajectory information \cite{Wang2023}.   Many methods aim to reconstruct a Boolean network model which combines simple parameterization and expressive dynamics.   Chen et al. used a genetic algorithm to reconstructed a Probabilistic Boolean Network from scRNA-Seq data, a model that accommodates probabilistic transitions in a Boolean network model \cite{Chen2014}, while CANTATA \cite{Kestler2022} and and SAILoR \cite{Moskon2024} use evolutionary strategies to find a best fit to discretized data.  Similar approaches that reconstruct a Boolean network model include ATEN  \cite{Shi2020}, which constructs an And/Or tree ensemble for representing the network interactions, RE:IN \cite{Yordanov2016}, which enumerates models that agree with experimental observations, and ASKeD-BN \cite{Smail2021} which tries to satisfy constraints on network structure and dynamics. 

Probability theory, on which the great majority of network inference approaches are grounded, is a very effective tool for modeling experimental noise in the measurement of individual genes' expression.  However, a network is a computational model whose output, which models a multidimensional experimental dataset, often corresponds to highly complex computation.  For example, there is a very high, nonlinear dependency between consecutive time points, which typically introduces more variation between datasets than that introduced by noise.  In order to focus on inferring the observed computation, a methodology based on a computational definition of randomness is required \cite{Kolmogorov1968}.  After the data is binarized to convey active/inactive gene states, the best network model is one that matches as many observed data bits as possible while at the same time has an encoding as short as possible.  Since the number of models that have a shorter encoding than a given sequence of bits is exponentially smaller in the size of the difference, a large difference corresponds to a non-random match .  The length of the encoding of the model is the logarithm base 2 of the number of possible models, and so the objective for optimization is the sum of mismatch bits plus the logarithm base 2 of the number of possible models.  We have previously introduced an algorithm that finds  a solution based on this criterion for a deterministic model \citep{Karlebach2023}.  However, our previous work did not address non-determinism in the computation.  Non-determinism is especially important for biological networks, where factors that we cannot model as part of the network often exist and create non-deterministic behavior.  Our goal in this work is to extend the framework described in  \citep{Karlebach2023} to apply to on-deterministic computations, or more specifically to datasets that reflect highly complex computations which include non-deterministic steps.  We will define an objective criterion for inferring a computational model from non-deterministic computational output and present an algorithm for finding an optimal solution with respect to this inference criterion, remaining within the framework of algorithmic information.  The model that we will use for this purpose is the Boolean network \cite{Fox2024,Samaga2009,Schwab2017}.  \newline

\section{Materials and Methods}\label{sec2}
\subsection{Network Inference}
\begin{figure}[t]
    \includegraphics[width=1.2\textwidth]{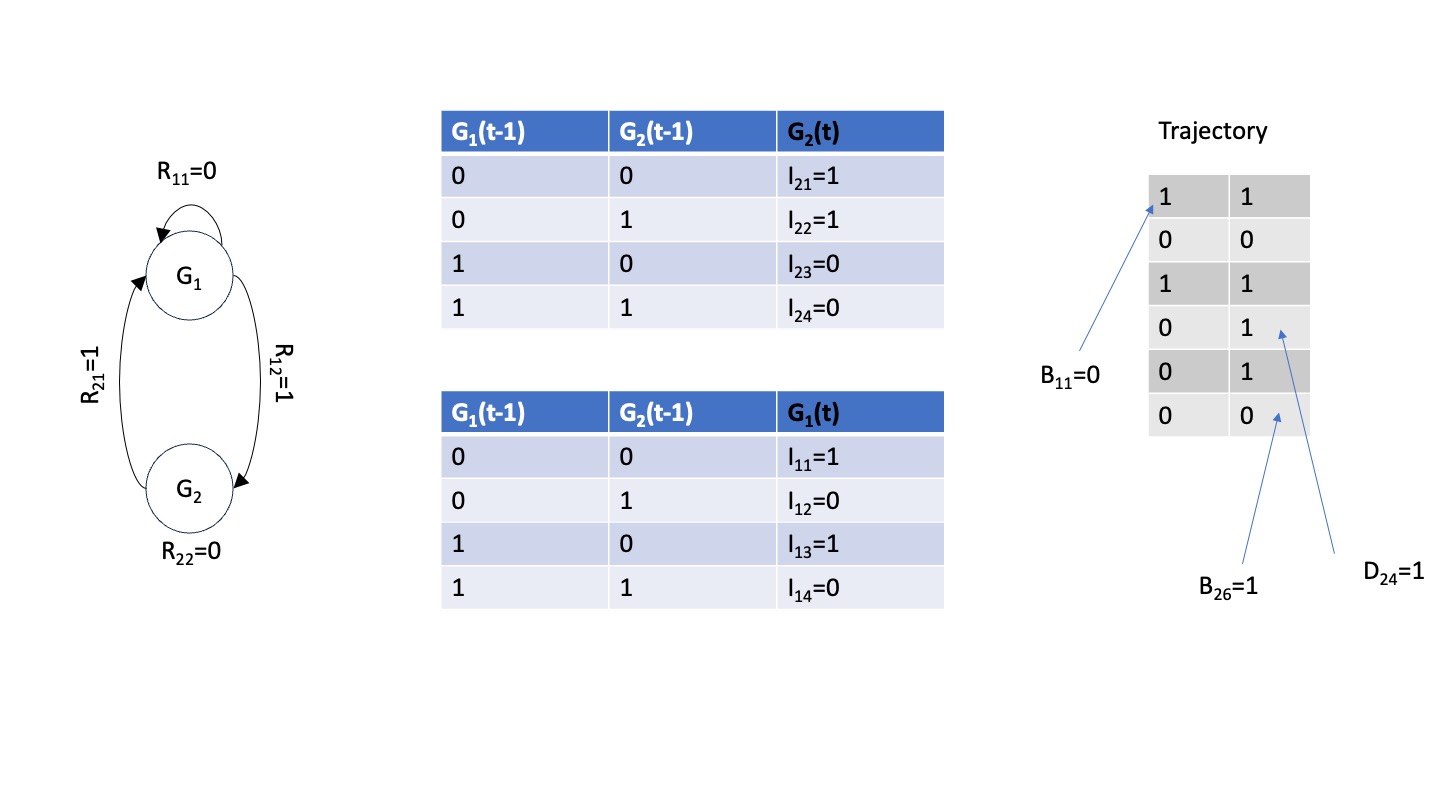}
     \caption{Illustration of a possible solution to the ILP formulation and the role of different variables in it.  The leftmost frame shows the nodes (genes), potential edges, the R variable corresponding to each edge and their value in the illustrated solution.  The middle frame shows the logic functions and the outputs selected for them by the solution.  These outputs define Boolean functions with a single input, in agreement with the R variable values. The rightmost frame shows the input trajectory and some of the values of B and D variables that ensure consistency between the selected edges, logic functions and inferred trajectory.}
     \label{fig:example}
\end{figure}

Discussing the inference of a Boolean network model from gene expression data requires some basic terminology.  The term \textit{state} (of a biological cell) refers to a vector $v \in \{0,1\}^N$.  Every gene is assigned a Boolean value, also known as the gene's level, in every state.  A Boolean network can be viewed as a function that maps a Boolean state to another, potentially identical, state.  That is, $BN: \{0,1\}^N \to \{0,1\}^N $.  For a given network, a sequence of states such that each state maps to its successive state in the sequence is called a \textit{trajectory}.   Hence a trajectory can also be represented as a matrix, where each column is a state.  The set of a network's possible trajectories contains distinct sequences of the form $v ,BN(v), BN(BN(v))...BN^M(v) \quad  \forall   v \in  \{0,1\}^N, M \in \mathbbm{N}$  A trajectory whose states are repeated indefinitely by the network is known as an \textit{attractor}.   Hence an attractor $\alpha$ satisfies $ \forall v \in \alpha , M \in \mathbbm{N}:  BN^M(v) \in \alpha$  An attractor consisting of a single distinct state is referred to as a \textit{steady state}.  Formally, a steady state is a state v such that $\forall M \in \mathbbm{N} : v = BN^M(v) $.  The trajectories and steady states generated by a network are known as the network \textit{dynamics}.  By definition, steady states and trajectories correspond to a specific network model $BN$.  Every such model is describe by a pair $(G(V,E),L)$, where $G(V,E)$ is a directed graph with a set of nodes (genes) $V$ and set of edges(regulatory relations) $E$, and $L$ is a set of Boolean functions $L=\{l_{v_i} :  2^{\lvert(u_i,v_i): (u_i,v_i) \in E \rvert} \to \{0,1\}\}$.  These are also known as the logic functions, and the genes whose Boolean values constitute their inputs are also known as regulators.  To compute $BN(v)$ for a state v, one computes $(l_{v_1}(u_1:(u_1,v_1) \in E),l_{v_2}(u_2:(u_2,v_2) \in E),...,l_{v_N}(u_N:(u_N,v_N) \in E) )$. Every function $l_{v_i}$ represents the effect of the biological regulation of gene $v_i$ by its set of regulators.  
A dataset is a set of trajectories and steady states, i.e matrices $T \in \{0,1\}^{N \times M}$ and vectors $S \in \{0,1\}^N$, possibly affected by noise, in which case states do not perfectly match the mapping defined by the network because some gene levels are incorrectly measured (\textit{noise bits}). That is, in a dataset for which the optimal solution consists of a network $BN$, it is possible that  $\exists i \in \{1,..M\} : T_i \ne BN(T_{i-1})$, where $T_i , T_{i-1}$ are column vectors of T. For a given network $BN$ and a trajectory $T_{M \times N}$, the number of noise bits is defined as $min_{v \in \{0,1\}^N} \lvert (i,j): T_{i,j} \ne  BN^j(v)_i \rvert$.  In words, given all possible initial states, the number of noise bits is the minimal number of differences between corresponding entries of the matrix representing the observed trajectory T and any trajectory matrix of the same size that BN can generate, i.e. over all possible initial states.  Intuitively, the noise is defined by the remaining unexplained bits after finding a best fit of the model to the data.  For multiple observed trajectories and steady states, the number of noise bits is the sum of noise bits of each individual trajectory and steady state.  
\newline
In algorithmic information theory \cite{Kolmogorov1968}, when considering deterministic Turing machines with an output alphabet of {0,1}, a machine with an encoding much shorter than an output string that generates that string constitutes a non-random solution, because the number of machines halves with every bit removed from the encoding.  Conversely, if the encoding of the shortest Turing machine that outputs a string is about the length of the output string, the string is assumed to be random, because every Turing machine of that encoding length could generate one of the possible output strings of that length, and so every string would have a Turing machine that outputs it.  A machine with encoding length not much longer than the string itself can be easily  formed by printing a hard-coded string(consider a computer program with a single print command).  We now need to adopt these arguments for Boolean networks and the presence of noise. \newline
The length of the binary encoding of a network is  the logarithm base 2 of the number of networks that we could form if we permuted the set of regulators of each gene and the outputs of the functions $l_i$. This is approximately $log_2(\prod_{i \in {1,2,..M}}{ \binom{n}{\lvert(u_j,v_i): (u_j,v_i) \in E \rvert}  \cdot 2^{\lvert(u_j,v_i): (u_j,v_i) \in E \rvert} } )=\sum_{i \in {1,2,..M}} { log_2(\binom{n}{\lvert(u_j,v_i): (u_j,v_i) \in E \rvert}) +\lvert(u_j,v_i): (u_j,v_i) \in E \rvert}$.   Intuitively, each such network generates a unique set of trajectories, and thus the number of networks that we can generate s proportional to the number of observed datasets that we can fit.  We should also point out that the number of \textit{non-redundant} Boolean functions of $k$ inputs, i.e. those where none of the function of k inputs are equivalent to functions with a smaller number of inputs, is slightly smaller than $2^k$, but for simplicity of explanation we will not use the longer expression in this text.  Our implementation of the method uses the number of non-redundant Boolean functions in order to calculate the length of encoding (see Data Availability).  Denote the number of noise bits as $\nu$.  Then the number of bits that are matched by the solution is $L-\nu$, where $L$ is the number of bits in the data.  If the length of the encoding is $\epsilon$, and every network can generate $T$ trajectories, the ratio between the number of possible bits that can fitted by a solution of the same size and the number of bit strings that can be observed in a fit of the same length is the order of $\frac{T \cdot 2^\epsilon}{2^{L-\nu}}= T \cdot 2^\frac{\epsilon+\nu}{L} $ By minimizing $\nu +\epsilon$, we obtain a solution with the smallest possible ratio, that is, a network (possibly empty) that is the most likely candidate to have generated the data and the locations of the noise bits.  At this point it is worth pointing out that as the sum $\nu+\epsilon$ becomes smaller, the number of datasets that will have a solution of that value becomes vanishingly small.   Furthermore, when a solution is found, it will also be the optimal solution for every assignment of Boolean values to the noise bits.  This can be easily shown but assuming towards contradiction that after flipping one noise bit the network BN is no longer optimal for the modified trajectory.  That means that there is another solution, such that the sum of network bits and noise bits on the modified dataset is less than this sum for BN.  But then if we flip the bit back to its original value, we added one noise bit to the solution defined by BN and at most one noise bit to the new optimal solution.  So the new solution is also better than BN for the original dataset as well - a contradiction to the optimality of BN.  This argument can trivially be extended to any number of bits.  So every bit of noise in the solution halves the number of trajectories that are not explained by the solution.   \newline   To add nondeterminism to these considerations, we define $l_i(t)$ as the output of the function $l_i$ at state number $t$ in the trajectory, where states are numbered by integers in order of occurrence.  If we allow $l_i(t)$ to remain equal to $l(t-1)$ and ignore $l_i$'s inputs at time t-1, we have introduced nondeterminism.  Intuitively, every time the network is allowed to use non-determinism, the number of computational paths doubles - half of them proceed without a non-deterministic transition and half of them with one.  Now suppose that every such "deferred change" in the function's output had a cost of 1 in the solution, i.e. the same cost as a noise bit, and consider a solution that uses one such non-deterministic transition for function $l_i$ at time point $t$.  We want to show that, just like with noise bits, this necessarily means that the number of trajectories that are explained by this solution $S$ is doubled.  We take the observed trajectory and replace its suffix, i.e. all states after time t, with states that match exactly the deterministic transitions of the network from the optimal solution, and thereby construct a new trajectory $T'$ from the original trajectory $T$.  Flipping the bits of $T'$ to match the bits of $T$ will flip at least one bit in the suffix, for otherwise a non-deterministic transition would not have been selected for the optimal solution at time $t$.   From the construction of $T'$, these different bits must be in the suffix, i.e. the states after time $t$.  Now suppose that there is a cheaper solution $S'$ that is inferred from $T'$.  Consider the trajectory $T_{S'}$ that $S'$ generates in its fit to $T'$.  If we flip the bits in $T'$ back to obtain $T$, some of them must introduce noise bits into $T_{S'}'$, for otherwise $S'$ would have been an optimal solution for $T$.  Denote this number of noise bits as $\nu'$.  Furthermore, the cost of $S'$ on the prefix of the trajectory $T'$ (and $T$, which has an identical prefix) must be at least $\nu'$ plus the cost of $S$ (which is 0 for the suffix), for the same reason.  But from this it follows that the cost of $S'$ on $T'$ is exactly $S$, in contradiction to the assumption that it is a cheaper solution. \newline
We now describe an algorithm for finding an optimal solution based on the optimality criterion that we described.  That is, the cost of every network encoding bit, noise bit and asynchronous/non-deterministic transition is 1.  We have previously described a solution based on Integer Linear Programming for the synchronous/deterministic case \citep{Karlebach2023}.  There are many similarities between the algorithms for the synchronous and asynchronous cases, as the models themselves share many similarities.  The differences and similarities will be stressed when describing the solution for the latter.  A gene expression dataset is a N $\times$ M matrix, where N corresponds to the number of genes whose expression level was measured, and M corresponds to the number of experiments.  The entry at indices i,j contains a Boolean value that is equal to 1 if the gene is in the active state, and otherwise equal to 0.  Initially, we assume that the order between states is known for every trajectory.  This is the case for biological datasets in which the cells have been synchronized or in which the cells are at steady state.  We will also assume that a set of plausible regulators has been determined for each gene.  This can be achieved using other experimental technologies like ChIP-Seq \cite{Mathelier2013}, or by keeping regulators whose levels correlate with their target's.  From this initial set, the modeler want to choose the optimal subset, including the logic tables by which the regulators determine the state of the target.  For mapping biological measurements to Boolean values we will use existing methods \cite{Hofensitz2012,Glaz2001,Shmulevich2002,Berestovsky2013}, and in this section we assume that all gene values are already Boolean.\newline
Since the methodology that we present is based on the concept of Kolmogorov complexity \cite{Kolmogorov1968}, one has to note that datasets may only contain any trajectory or steady state once.  This is because while non-determinism can occur in trajectories, the model and the inference algorithm do not make use of the concept of probability.  Intuitively, if we run a computer program with the same input twice (including the same seeds for pseudo-random number generators) we do not obtain new information about the program from observing the same output in both runs.  The variables of the ILP problem are denoted by uppercase English letters.  A $B$ variable is defined for every measurement, i.e. an entry in the expression matrix that describes a gene and its activity at a given state, and is equal to 1 if there is a mismatch between the observed value of the gene at that measurement and the value that the model assigns it.  An $I$ variable is defined for every combination of regulator values, and is equal to 1 if the state of the target gene is set to 1 for that combination, and otherwise it is equal to 0.  That is, the value assigned to the $I$ variable in the optimization determines the output of the logic function for a specific input. An $R$ variable is defined for every potential (regulator,target) pair.  It is equal to 1 if the regulator is chosen in the optimal solution, and otherwise to 0.  A $V$ variable is defined for every gene and possible number of regulators for that gene, and is equal to 1 if the gene has at least that number of regulators, and otherwise it is equal to 0.  Later in this section we will also define a $D$ variable, which will allow us to implement asynchronous dynamics.   Figure \ref{fig:example} illustrates the roles of different variables in a possible solution for a 2-gene network.  
Using these variables, we first describe the constraints of the model, and then the objective function:
Let $C_{i,j}$ denote the observed Boolean value of gene $i$ at experiment $j$.  The corresponding $B$ variable is $B_{g_i,j}$, and it is equal to 1 if the value of gene $g_i$ in experiment $j$ does not match the model's assignment, and otherwise 0.  If $j$ is the index of a steady state in the data, $g_{k+1}$ is a gene with regulators $g_1,g_2,...,g_k$, we go over every possible combination of values for these regulators $(w_1,w_2,...,w_k)$    ,    $w_j \in \{0,1\}$ and for each combination add the following constraint:
\begin{equation}
 \sum_{r=1}^{k} ( C_{r,j} \cdot (w_r+(1-2 \cdot w_r) \cdot B_{g_r,j})    
 \end{equation}
 \begin{equation}
  + (1-C_{r,j}) \cdot ((1-w_r)+ (2\cdot w_r-1) \cdot B_{g_r,j}))  \nonumber
 \end{equation}
\begin{equation} 
  + C_{k+1,j}\cdot B_{g_{k+1},j}+(1-C_{k+1,j})\cdot (1-B_{g_{k+1},j})    \nonumber
\end{equation}
\begin{equation} 
  < (2 -I(w_1,w_2,...,w_k))\cdot(k+1)    \nonumber
\end{equation}

where  $I(w_1,w_2,...,w_k)$ is the output of the Boolean function that determines the value of $g_{k+1}$.
This constraint means that if the output variable $I(w_1,w_2,...,w_k)$ was set to 1, whenever the combination $w_1,w_2,...,w_k$ appears, the output (the value of $g_{k+1}$) must be 1.  If the data contains trajectories, the observed values of the target gene and the corresponding 0/1 IP variables will be taken from the subsequent time point, at which the regulation is expected to take effect if the model is synchronous.
Similarly, we add the following constraint to account for the case where $I(w_1,w_2,...,w_k)$ is set to 0,:
\begin{equation}
 \sum_{r=1}^{k} ( C_{r,j} \cdot (w_r+(1- 2\cdot w_r) \cdot B_{g_r,j})     
 \end{equation}
 \begin{equation}
  + (1-C_{rj}) \cdot ((1-w_r)+ (2\cdot w_r-1) \cdot B_{g_r,j}))  \nonumber    
 \end{equation}
\begin{equation} 
  + C_{k+1,j}\cdot(1- B_{g_{k+1},j})+(1-C_{k+1,j})\cdot B_{g_{k+1},j}   \nonumber
\end{equation}
\begin{equation} 
  < (I(w_1,w_2,...,w_k)+1)\cdot(k+1)    \nonumber
\end{equation}
\linebreak
Next, for every gene $g_i$ and each one of its regulators $g_j$, we create a Boolean variable $R_{ij}$.  In other words, every potential regulator of gene $g_i$ is associated with an $R$ variable for that gene.  For every two different assignments of values to $g_i$'s regulators, i.e. inputs to the logic function that sets the value of $g_i$, the sum of $R$ variables of regulators which have different values in the two assignments is constrained to be greater than the differences between the $I$ variables that determine the outputs for these assignments. For example, with two regulators $R_1$ and $R_2$ and two assignments $0 1$ and $0 0$ to the variables, respectively, $R_2$ must be greater than $I_{01}-I_{00}$ and also greater than $I_{00}-I_{01}$.  If the outputs for these two assignments are different, only the change in $R_2$ can explain this difference,as $R_1$ has the same value in both assignments.   More generally, this constraint means that two different outputs can never occur for the exact combination of regulatory inputs, for otherwise the regulatory logic is not a function.
The $V_{ik}$ variable, which is defined for gene $i$ and every possible number of regulators $k$ of that gene, is constrained to be greater than the mean of the gene's R variables minus $\frac{0.5}{indegree(g_i)}$ if $k=1$ or $(\frac{i-1}{indegree(g_i)})$ if $k>1$, where $indegree(g_i)$ is the number of candidate regulators (or $R$ variables) of $g_i$.
To set the weights of variables in the objective to match the inference criterion, a weight of 1 is given to $B$ variables.  Now if $r$ regulators are chosen for a gene, all its $V$ variables $1..r$ will be set to 1.  Therefore, we set the weight of the first $V$ variable of the gene to be the logarithm (base 2) of the number of ways to choose a first regulator plus the $log_2$ of the number of logic tables possible for one regulator.  We then set the weight of the second $V$ variable of the gene to be the $log_2$ of the number of ways to choose a second regulator after the first one was already chosen plus the $log_2$ of the number of logic tables with two regulators, minus the $log_2$ of the number of logic tables with one regulator.  So the costs of encoding the logic tables cancel out by consecutive $V$ variables, while the cost of choosing the regulators is produced by the combination of all the $V$ that are set to 1.
If we denote the number of logic tables with $k$ regulators as $L_k$, the weight of the $k^{th}$ $V$ variable is set to $log_2(L_k)-log_2(L_{k-1})+log_2(\binom{N-k+1}{k})$ 

This part of the formulation coincides with the synchronous formulation.  We now adapt the 0/1 IP formulation to fit asynchronous dynamics.  We add a new type of variable called the $D$ variable.  This variable is defined for every constraint that involves the $B$ variables in a trajectory, as defined in (1) and (2).  It is added to the right hand side of the constraint, and therefore if it is equal to 1 it allows the output of the logic function to not agree with its inputs.  We further constrain the $D$ variable to be smaller than 1 minus the differences between the chosen value of target gene at the state at which the regulatory effect is taking effect and the previous state, i.e., the values selected for the gene by the model at these states.  The latter constraint only allows the output of the logic function to disagree with its input if the output does not change, i.e. if the regulatory update is not immediate.  Using the same notation as before, the additional constraints on the $D$ variable can be described as follows:

\begin{equation}
1-(C_{k+1,j+1}*(1-B_{g_{k+1,j+1}})+(1-C_{k+1,j+1})*B_{g_{k+1,j+1}}  \nonumber
\end{equation}
\begin{equation}
-(C_{k+1,j}*(1-B_{g_{k+1,j}})+(1-C_{k+1,j})*B_{g_{k+1,j}}))>=D   
\end{equation}

\begin{equation}
1-(C_{k+1,j}*(1-B_{g_{k+1,j}})+(1-C_{k+1,j})*B_{g_{k+1,j}}			\nonumber
\end{equation}
\begin{equation}
-(C_{k+1,j+1}*(1-B_{g_{k+1,j+1}})+(1-C_{k+1,j+1})*B_{g_{k+1,j+1}}))>=D   
\end{equation}

Finally, for each target gene we constrain the first $D$ variable in each trajectory to be smaller or equal to the sum of the target's $R$ variables, such that the target would only be able to use the $D$ variables if it has at least one regulator assigned to it.  We set the weight of every $D$ variable to 1 in the objective function.  The value of the objective function is then the sum of the weights of variables that are set to 1 in the solution, i.e.: 
\begin{equation}
 \sum_{i=1,j=1}^{N,M} (B_{g_i,j}+W( V_{g_i,j}) \cdot V_{g_i,j}+D_{g_i,j})    
 \end{equation}
where W is a function that maps between a V variable and the weight assigned to it.  This concludes the 0/1 Integer Programming formulation.  A concise formulation is also provided in the supplement.  An object-oriented implementation in Python also provides a similar specification of the ILP problem.  \newline

Powerful solvers like Gurobi \citep{gurobi} have dramatically improved our ability to solve 0/1 Integer Programming problems.  Custom heuristics can be integrated with the solver to improve performance.  We now describe such heuristics.

Perhaps the simplest heuristic for matching a given network structure to a trajectory is to perform a single pass over the trajectory, state by state starting from the first state, and to record every input-output pair observed as long as it does not conflict with pairs observed before it.  When a conflict occurs, the value of the target gene is flipped to match the output that was previously observed.  This way a consistent logic for the network structure is found, at the cost of introducing noise bits into the solution.  Inconsistencies are bits in the data that do not agree with a given network logic and its input.  This can be a result of either noise or a wrong/incomplete network structure and logic.   We shall refer to this simple heuristic as the single-pass heuristic.  A more sophisticated approach was suggested by Karlebach and Robinson \citep{Karlebach2023}, and can be applied to an expression data set composed of either steady states or equal-length trajectories:

\begin{algorithm}
\caption{Heuristic Search Algorithm}\label{heur}

1. Choose a set of regulators.  This set forms a network structure (without defining logic functions). \newline
2. If the set of states is composed of a single steady state, return it as a solution. \newline
3. If the set consists of a single trajectory, solve any inconsistencies using the single-pass heuristic, and return it as a solution. \newline
4. If the size of the set of steady states or trajectories is larger than 1 but the set is consistent with the regulators, i.e. we can choose logic functions such that no noise bits will be introduced, return that set of states as a solution.  Since some regulators (edges in the network structure) may be redundant, i.e. logic functions will have inputs that do not affect the output, it is possible to remove some or all of them by backward elimination. \newline
5. Otherwise, the set of states does not represent a trajectory that is generated by the network structure, i.e. at least one of these trajectories cannot be reproduced by the network structure with any logic function assignment.  In this case, cluster the states and round the cluster centers into Boolean vectors, then solve the problem recursively for the cluster centers.  The recursive call returns a set of consistent states S.  For every state in the original set, choose its closest neighbor in S, and flip its values one by one to match the neighbor's values until all inconsistencies with states in S have been resolved, or until it is equal to the neighbor, which is already consistent.  At that point add it to S so it can be compared to states that have not been made consistent yet when those are added.  At the end of the process, return S excluding the cluster centers.  \newline

\end{algorithm}

\begin{algorithm}
\begin{algorithmic}
\caption*{Heuristic Search pseudocode}\label{heurp}
\State Heuristic\_Search(Edges,States):

\If{$length(States) = 1$}
	\If{$type(States) = trajectory$}
	\If{$not \  logic\_is\_consistent(Edges,States)$}
		\State States=single\_pass\_heuristic(Edges,States)
	\EndIf
		\State return (keep\_required\_edges(Edges,States),States)
	\ElsIf {$type(States)==steady\_state$}
		\State return (Edges,States)
	\EndIf
\EndIf	
\If {$logic\_is\_consistent(Edges,States)$}

	\State return (keep\_required\_edges(Edges,States),States)
\EndIf	
\State $new\_states \gets cluster\_centers(States)$
\State  $new\_edges,new\_states \gets Heuristic\_Search(Edges,new\_states)$
\For {$state \in States$}

	\State $closest\_state \gets find\_min\_hamming\_dist\_neighbor(state,new\_states)$
	
	\State $i \gets 0$
	
	\While {$not \  logic\_is\_consistent\_logic(new\_edges, union(new\_states,state))$}
	
		\State $state[i] \gets closest\_state[i]$
		
		\State $i \gets i+1$
	\EndWhile
\EndFor
		
\State $return (keep\_required\_edges(new\_edges,States),States)$
\end{algorithmic}	
\end{algorithm}

Clustering refers to dividing the states into distinct groups such that states within each group are more similar to each other than to states in other groups, based on a given similarity measure.  The cluster centers are the vectors that represent the average vector in the cluster, and are typically determined by averaging each vector entry separately.  Since this averaging may not result in a Boolean value, typically we would map values above 0.5 to a Boolean 1 and other values to 0, though this mapping can follow a more sophisticated procedure that may affect the effectiveness of the heuristic.  The set of regulators that forms the structure in step 1 can be chosen from the current LP solution, for example all regulators which correspond to an $R$ variable with value of at least 0.5.  If the dataset contains both steady states and trajectories, then the recursive heuristic can be run for the steady states, and then the resulting logic can be used to remove inconsistencies from the trajectories using the single-pass heuristic.  If trajectories have different lengths, equal-sized contiguous subsequences of trajectories can be solved by the recursive heuristic, and the remaining inconsistencies then resolved by the single-pass heuristic.  Care should be taken that clustering of these subsequences is biologically meaningful, for otherwise poor solutions may be result due to their incompatibility. A pseudocode for the heuristic is provided on page ~\pageref{heurp}. \newline

It remains to adapt the heuristic to allow for asynchronous dynamics.  In the adapted version, if a gene's value does not match the output expected by the values of its regulators, but it is consistent with the value of the gene in the previous time point, then it is no longer flagged as an inconsistency.  Additionally, when fixing inconsistencies by performing a pass over trajectories and building a set of logic functions, functions are only  updated when their target genes change their values between consecutive time steps.  With these changes, the heuristic can be applied to asynchronous trajectories, or a combination of steady states and such trajectories. For ease of reference we will refer to this procedure as MEDSI (Minimum Edit Distance from a State of Ignorance). \newline

It should be noted that the heuristic described in \ref{heur} is in fact a family of heuristics.  The order in which Boolean values are flipped in step 4 does not have to be arbitrary, and different criteria for this order will result in different solutions.  Similarly, there are multiple ways to select a set of regulators, and various ways to cluster the states.  Different choices will result in different heuristics.  For example, one similarity measure may cluster states that differ not only by noise bits, whereas another may assign these states to different clusters.  This will affect the quality of the heuristic solution and thus the convergence to an optimal solution.
\subsection{Computational Complexity}
For a Boolean network with N genes, where each gene has K potential regulators,  there are $2^{K\cdot N}$ possible structures, and for each such structure on average about $2^{N \cdot K/2}$ logics.  For N=10 genes and K=3, this gives ~$35 \cdot 10^{12}$ possible networks that differ in topology, logic or both.  Since this makes brute-force searches impractical, in this section we discuss the relationship between input size and running time.
In \cite{Karlebach2012} the Vertex Cover problem is reduced to the network inference problem by mapping each vertex to a regulator, and each edge that it covers to a target gene of that regulator.  The network structure is assumed to be fixed, and the noise bits (represented by the $B$ variables) are minimized.  Since the resulting dataset consists of steady states, it applies to both synchronous and asynchronous networks.  
Let us now consider the same reduction, but when the network is created, each edge is mapped to multiple nodes, with the same set of regulators.  I.e, if $e=(u,v)$ is an edge in the graph, in the network that it is reduced to we add multiple copies of a target gene $g_e$, all having the same set of regulators $\{g_u,g_v\}$.  The regulators $\{g_u,g_v\}$ have no regulators of their own, as in the original reduction.  Each such target gene $g_e$ will have a Boolean value of 0 in one steady state and a Boolean value of 1 in another steady state, as in the original reduction.  The regulators have the same Boolean values in both steady states.  Consequently, either the Boolean values of the target or the Boolean values of the regulators need to be changed, i.e. to be considered noise bits.  If we fix the structure, meaning that all the regulators are selected for each target gene, then we don't have to consider solutions where regulators are not selected.  Similarly, since each target gene has 2 regulators, the cost of adding them to the solution is constant, and therefore by creating a number of copies of every target gene that is polynomial in the input size, we can in effect fix the structure, because introducing noise in all the target's copies costs more than assigning it two regulators and introducing a noise bit in one of them in one of the steady states. 
Hence, the problem described in this paper is NP-Complete as well - allowing a choice of network structure and logic does not reduce the computational complexity of the problem.  We observed that in practice, higher levels of noise will result in longer running times, whereas for noiseless datasets the minimal network can be found relatively fast.  Interestingly however, the dataset constructed by the reduction in \cite{Karlebach2012} can be modified not such that the solution does not require the introduction of noise.  That is, once the network structure has been selected, none of the measured gene values needs to be changed.  This can be done by modifying the reduction - we set the Boolean value of every regulator to 1 in one steady state and to 0 in the other.  Now if a regulator is chosen for the solution, no noise bits are required, since both the regulator and the target change their values between the two steady states.  This is again a better solution than introducing noise bits, because there are multiple copies of each target gene and they all share the same pair of regulators.  By choosing every graph nodes $u$ that correspond to a selected regulator $g_u$, we ensure that every edge in the original graph is covered.  The solution is minimal, for otherwise there would have been a better solution for the network inference problem.  Hence, assuming $P \ne NP$, in the worst case the structure inference alone, in the absence of noise, will require exponential time.  Since the computation of the optimal network is performed in-silico, we argue that it constitute a worthwhile utilization of resources.   The Simulation section describes the gains in accuracy that are obtained by solving this problem.  It is also important to notice that when using an ILP solver, a good part of the computation is devoted to proving the optimality of a solution after the optimal solution has been found, and often an optimal solution can be found even if a time limit is imposed.
One should note that if the number of regulators considered for a gene is order of the number of genes, the ILP algorithm will have exponential complexity.  Providing a set of candidate regulators for each gene in the order of the logarithm of the number of genes avoids this complication.

\section{Results}\label{sec3}

\subsection{Novel Network Inference from Expression Data}
\begin{figure}[t]
    \includegraphics[width=0.9\textwidth]{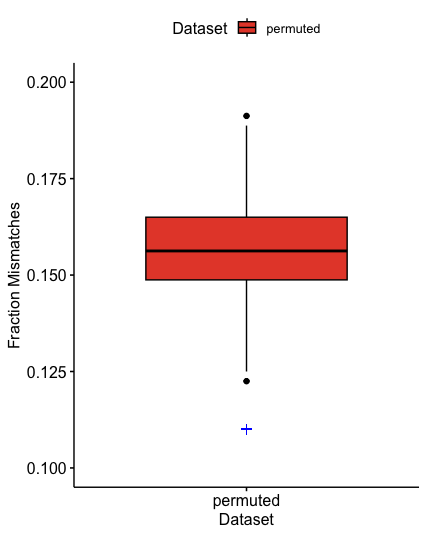}
     \caption{Distribution of the fraction of mismatches when the model that was inferred using the test set is fitted to 1,000 randomly permuted datasets.  The fraction of mismatches when fitting the inferred model to the real test set is marked by a blue plus sign.  In all the permuted datasets the fraction of mismatches was higher, which corresponds to a p-value of 0.001.}
     \label{fig:permtest}
\end{figure}

We start by constructing a network of hypoxia \cite{Kwast1999} response in yeast.   Since the ground truth is not known, we will evaluate the network as follows:  split the data into two equal-sized parts - training set and test set.  Infer the model on the training set.  Then, keep the inferred topology and logic functions fixed, and compute the number of mismatches in the best fit to the test set.  To compare the resulting fit to random fit, permute the test set 1,000 times and fit the inferred model to each permuted test set.  If the fit to the real test set is better (less mismatches) than the permuted test sets, the model is predictive on unobserved data, i.e. the test set.  In order to construct a network of transcriptional interactions, we extracted transcriptional interactions of genes regulate the transcriptional response to stress from Yeastract \cite{Monteiro2019}.  We obtained from GEO \cite the dataset GSE115171, which includes time-series of yeast with various deletions under hypoxia.  In order to account for deletions in the modeling, we set the values of a deleted gene in the relevant experiment to 0, its $B$ variable in the experiment is set to constant 0 and the regulatory constraints ((1) and (2)) are removed for that target gene in the relevant experiment/trajectory.  Binarization of the RNA-Sequencing data is performed using the BASCA method \cite{Hofensitz2012}, applied to the log-transformed, library-size-normalized RNA-Seq counts.   The fraction of mismatches when fitting the inferred model to the test set was 0.11.  The lowest fraction of mismatches when fitting the model to permuted tests sets was 0.1225.  In terms of null-hypothesis significance testing, the p-value of observing this result by chance is less than 0.001.  Figure \ref{fig:permtest} illustrates this result, where the boxplot shows the distribution for permuted datasets and the blue plus sign shows the fraction of mismatches for the real test set.  Since the test set was not used in inference but was predicted effectively by the model, and since it consists of trajectories, we conclude that the inferred model captures not only the topology but the logic and asynchronous dynamics of the real transcription-regulatory network.

\subsection{Simulation using a Literature-derived Network}
To test our method using a gene regulatory network obtained from the literature, we used the cell cycle model of Cho et al. \cite{Cho}.  Using the R package Boolnet \cite{BoolNet}, we generated 100 time points divided into 4 trajectories, i.e. 4 trajectories of 25 time points each, including both synchronous and asynchronous transitions and 15\% noise, i.e. the probability of a bit flip in the generation of the trajectories is 0.15.  We also added a random false regulator to each gene, i.e. candidate regulator that had no effect on the generated trajectories.  Since we know the ground truth, we can compare our methods to other approaches.  The MCC \cite{Matthews1975} obtained by our approach (denoted as MEDSI) was 0.57, and none of the false regulators were selected.  For the tools BestFit\cite{Lhdesmki2003}, ATEN \cite{Shi2020} and CANTATA \cite{Kestler2022} the MCC was 0.22, -0.14, and 0.02 respectively.  This experiment shows that for a network structure and logic that were curated by experts from the literature, we can infer biological regulation reliably, and exclude regulations that are not reflected in the data.  Asynchronicity in this case may be a consequence of biological mechanisms that are not part of the gene regulatory network - for example, G$_1$ inhibitors \cite{Cho}.
\newline

\subsection{Simulation using Random Networks}
In order to evaluate the accuracy of our inference algorithm, we generated random networks with various types of topologies:  fixed, scale-free and homogenous, as implemented in the R package BoolNet \cite{BoolNet}.  In addition to the real edges generated by BoolNet we added a random false regulators to every gene's list of regulators.  Datasets included both steady-states and trajectories, or only trajectories if the tool did not support steady-state data.  The trajectories included asynchronous updates and noise, and steady states included noise but by definition cannot include asynchronous updates.  The other tools shown in the comparison all reconstruct a Boolean network model and have a stable publicly available implementation, and include BestFit \cite{Lhdesmki2003}, ATEN  \cite{Shi2020} and CANTATA \cite{Kestler2022}.   The ATEN R package comes with its own function to generate time-series which we use for generating its input for each one of the test datasets, with parameters matching the properties of each dataset.  Since ATEN does not provide an option to choose from a subset of edges, we multiply the number of false positives that it makes by a factor of $ \frac{|S_N|}{(N-1) \cdot N} $, where $N-1$ is the total number of potential regulators (the number of genes minus 1) and $S_N$ is the subset of regulators that is considered in the dataset.  For example, if there are four potential regulators and two are considered in the dataset, every two false positives that ATEN makes is counted as one.  CANTATA does accept an initial network structure, but can also add other edges to it.  Therefore, we do not count any other edges that CANTATA adds as false positives. The total number of network states for each simulated dataset was set to 500, and noise levels and network topology varied between different simulated datasets.  Trajectories were extended until steady state for tools that supported varied length, and otherwise held fixed.  MEDSI performed the same whether trajectory length was fixed or varied.

\begin{figure}[t]
    \includegraphics[width=.3\textwidth]{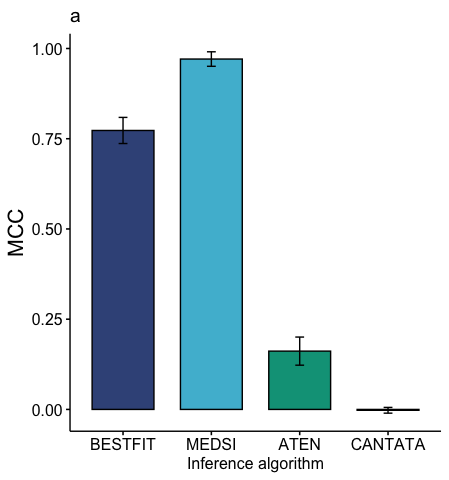}
    \includegraphics[width=.3\textwidth]{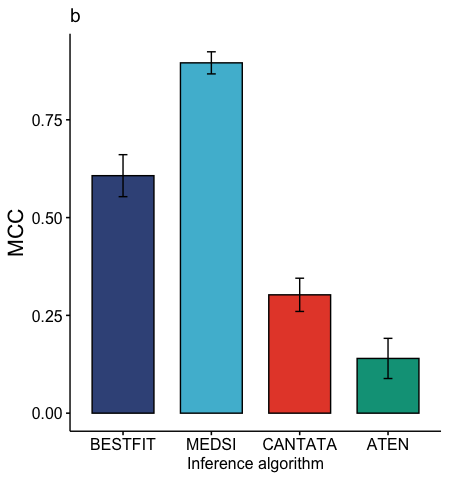}
    \includegraphics[width=.3\textwidth]{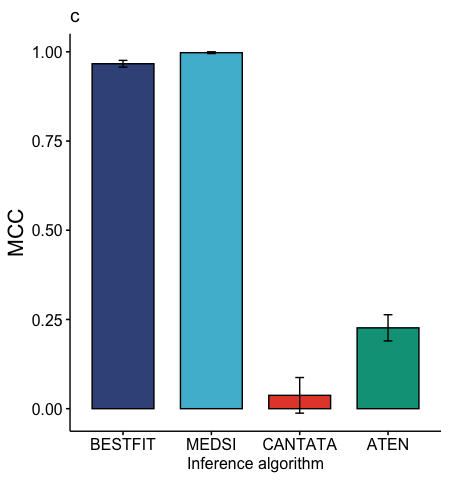}
   
    \includegraphics[width=.3\textwidth]{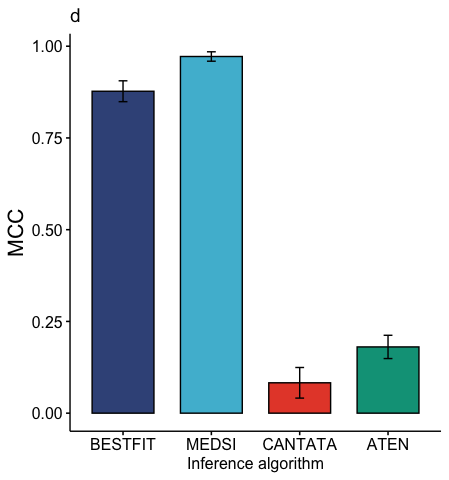}
    \includegraphics[width=.3\textwidth]{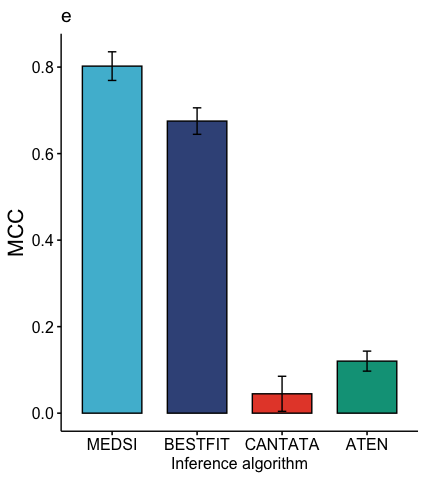}
    \includegraphics[width=.3\textwidth]{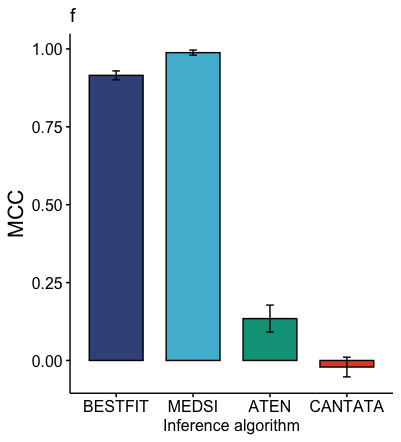}

     \caption{Mathews correlation coeffcient for the method described in this manuscript (MEDSI) and other tools, computed for 6 different simulations that vary in network topology and datasets properties.  The vertical bars correspond to the standard error of the mean.  Frames a,b and c correspond to fixed, homogeneous and scale-free topologies, respectively, with noise level 0.1, while frames d-f correspond to a scale free topology with noise levels 0.15,0.15 and 0.05, respectively.}
     \label{fig:mcc}
\end{figure}

The Mathews correlation coefficient (MCC) \cite{Matthews1975} obtained by the different tools on the fixed, homogeneous and scale-free topology datasets is shown in figure \ref{fig:mcc} frames a,b and c, respectively,  where the y-axis corresponds to the mean MCC.  As can be seen in the figure, our approach (MEDSI) achieves the highest score.   For MEDSI, the lower score for the homogeneous topology was due to false-negatives - it chooses less edges, as the nature of the dynamic behavior requires more data to make safe inferences.  For BestFit, the lower score for the fixed and homogeneous topologies was due to a higher number of false-positives.  This indicates that in the scale-free topology where most genes have a single regulator changes in regulator levels are well-correlated with their targets, providing that trajectories contain distinct states, many samples are available and noise level is about 10\%, but otherwise the accuracy of methods that rely on correlation quickly drops.  To consolidate this observation, we generated three additional collections of networks and datasets where the topology is always scale-free but the dataset properties vary: for the first dataset (figure \ref{fig:mcc}d) trajectory length is fixed at 10 and noise level increase to 0.15, in the second dataset (figure \ref{fig:mcc}e) the noise level increases to 0.15 and it contains only 50 network states, and in the third dataset (figure \ref{fig:mcc}f) there are twice as many false regulators and the noise level is reduced to 0.05.  MEDSI again outperformed other methods, and the performance of BestFit dropped compared to its performance in the tests illustrated in figure \ref{fig:mcc}c.  Hence modeling the underlying computational process results in better predictions, while methods that rely on correlation will reach a number of wrong conclusions that depends on network topology and dataset properties.
\newline

\section{Conclusion}
In this work we presented a novel network-inference methodology that finds an optimal solution with respect to minimizing spurious fits, and can account for asynchronous updates in network dynamics.  The inferential procedure balances model encoding length, noise bits and use of asynchronicity, and the objective of inference is based on the concept of Kolmogorov complexity.  Our methodology is implemented in software and publicly available for the community.  It provides an accurate tool for researchers of complex systems, in particular gene regulatory networks.  Due to the computational complexity of the problem, we introduced heuristic procedures that can enhance the efficiency of the exact ILP solution.   Based on the experiments presented in this work we believe that the algorithm is applicable to a broad range of high-throughput datasets, and can be used to understand the asynchronous dynamics of transcriptional regulation. Several objectives are left for future work:  First, additional heuristics should be developed and studied, as the problem it addresses is likely to present a variety of challenging instances.   Second, the topic of binarization of continuous or discrete data into Boolean values should be further pursued to obtain better understanding of current experimental technologies.  For example, an effective method for overcoming time-scale differences of different regulatory effects would result in more accurate binarization and consequently less noise bits.  Finally, combining different types of biological networks into a single model can provide broader insights into cellular function, and its optimal processing for use with the method described in this work should be studied.

\backmatter

\section*{Declarations}

\subsection{Competing Interests}
The authors declare that they have no competing of interests.  
\subsection{Data Availability}
An implementation of the method described in this work can be found at https://github.com/karleg/MEDSI
The data is used in this work is available publicly at the Gene Expression Omnibus at https://www.ncbi.nlm.nih.gov/geo/
\subsection{Ethics Declaration}
Not applicable.
\subsection{Consent to Publish Declaration}
Not applicable.
\subsection{Consent to Participate Declaration}
Not applicable.
\subsection{Author Contributions}
G.K. performed all the work described in this manuscript.
\subsection{Funding}
There is no funding to declare.
==========================================================

\bibliography{sn-bibliography}

\end{document}